# Small Energy Gap Revealed in CrBr$_3$ by Scanning Tunneling Spectroscopy

Dinesh Baral, Zhuangen Fu, Andrei S. Zadorozhnyi, Rabindra Dulal, Aaron Wang, Narendra Shrestha, Uppalaiah Erugu, Jinke Tang, Yuri Dahnovsky, Jifa Tian, & TeYu Chien*

*Department of Physics & Astronomy, University of Wyoming, Laramie, WY 82071*

CrBr$_3$ is a layered van der Waals material with magnetic ordering down to the 2D limit. For decades, based on optical measurements, it is believed that the energy gap of CrBr$_3$ is in the range of 1.68-2.1 eV. However, controversial results have indicated that the band gap of CrBr$_3$ is possibly smaller than that. An unambiguous determination of the energy gap is critical to the correct interpretations of the experimental results of CrBr$_3$. Here, we present the scanning tunneling microscopy and spectroscopy (STM/S) results of CrBr$_3$ thin and thick flakes exfoliated onto highly ordered pyrolytic graphite (HOPG) surfaces and density functional theory (DFT) calculations to reveal the small energy gap (peak-to-peak energy gap to be 0.57 eV $\pm$ 0.04 eV; or the onset signal energy gap to be 0.29 $\pm$ 0.05 eV from d$I$/d$V$ spectra). Atomic resolution topography images show the defect-free crystal structure and the d$I$/d$V$ spectra exhibit multiple peak features measured at 77 K. The conduction band – valence band peak pairs in the multi-peak d$I$/d$V$ spectrum agree very well with all reported optical transitions. STM topography images of mono- and bi-layer CrBr$_3$ flakes exhibit edge degradation due to short air exposure (~15 min) during sample transfer. The unambiguously determined small energy gap settles the controversy and is the key in better understanding CrBr$_3$ and similar materials.

---

*Corresponding author: tchien@uwyo.edu

Two-dimensional (2D) materials are state-of-the-art materials for the paramount goal of engineering in the 21st Century - minimizing electronic devices toward the atomic level [1–5]. Starting from 2004, the realization of monolayer graphene [1] led to the explosive explorations of 2D materials. This expands to the entire spectrum of material categories, ranging from insulators, semiconductors, semimetals, metals, to superconductors [1–5]. In 2017, two groups independently demonstrated stable magnetic ordering in 2D materials ($CrI_3$ and $Cr_2Ge_2Te_6$) [6,7], adding the 2D ferromagnetism as a new member of the 2D material family. Since then, 2D magnetism has been demonstrated in various van der Waals (vdW) materials, such as the more air stable $CrBr_3$ [8,9].

Magnetic [10–12] and optical [13–15] properties of $CrBr_3$ have been studied since 1960s; while the electronic properties were studied later [16,17]. Bulk $CrBr_3$ has a saturated magnetization of $\sim 3\mu_B$ per $Cr^{3+}$ ion [18–20] with a magnetic transition temperature of 32 K [21,22], and has been confirmed to have ferromagnetic ordering at its monolayer limit [8,9]. Optically, many features with distinguished energies were reported by various techniques, such as absorption [14,15,17,23,24], reflectance [25], photoluminescence (PL) [26] and Kerr rotations [14]. For example, $CrBr_3$ shows positive and negative Kerr rotation peaks at 23,500 cm$^{-1}$ (2.92 eV) and at 26,700 cm$^{-1}$ (3.29 eV), which are assigned to charge transfer (CT) transitions of electrons between Cr and Br atoms [14]. Absorption spectrum measured on $CrBr_3$ shows peaks corresponding to 13,500 cm$^{-1}$ (1.68 eV), 14,400 cm$^{-1}$ (1.79 eV), 17,500 cm$^{-1}$ (2.17 eV), 18,900 cm$^{-1}$ (2.36 eV) [17,23,24], 24,500 cm$^{-1}$ (3.04 eV) and 29,500 cm$^{-1}$ (3.66 eV) [14,15]. Among them, the lowest energy absorption, 1.68 eV, has been frequently quoted to be the energy gap of $CrBr_3$, and the consequent DFT calculations reproduced this energy gap. However, a recent PL measurement shows a peak with energy of 1.35 eV [26] assigned to *d-d* transition in CrBr$_3$, posing a controversy on the energy gap of CrBr$_3$.

CrBr$_3$ is known as a Mott insulator [27–31] with ferromagnetic ordering [32], where a large *d-d* Coulomb correlation energy $U$ (Coulomb repulsion energy between two electrons with opposite spin orientations on the same site) introduces an energy gap [33–36]. So far, the electronic properties of CrBr$_3$ were studied by photoemission spectroscopy [16], scanning tunneling spectroscopy [9,37], electrical transport [17], and density functional theory (DFT) calculations [20,25,38]. Despite the efforts, the understanding of the electronic properties is still not satisfactory. For example, the density of states (DOS) of CrBr$_3$ calculated by DFT using different setups can be very different [39–41]. In particular, comparisons with the experimental data, especially the optical features, are inconsistent. These inconsistencies posed an important question: *what is the value of the energy band gap of CrBr$_3$*? The answer to this question is critical to unambiguously explaining experimental data involving CrBr$_3$ and critical for accurately determining DFT calculation settings. For example, the different DFT calculation settings may influence the prediction of the strained induced anti-ferromagnetism vs ferromagnetism, similar to the case of CrI$_3$ [42,43].

The optical measurements and DFT calculations have inconsistent conclusions in energy gaps. Absorption spectra show a 1.68 eV [24] or 2.1 eV [17] energy gap whereas PL data shows a *d-d* transition [26,30] of 1.35 eV. DFT calculations reported energy gaps of CrBr$_3$ in a very wide range: 1.15 eV – 2.95 eV [20,30,31,38–41]. It was pointed out that the inclusion of the on-site Coulomb repulsion $U$ of Cr 3*d* electrons is crucial for the calculation of CrX$_3$ (X = Cl, Br, and I) and it is usually a semi-empirical parameter obtained by directly comparing physical properties [40]. The corresponding correlation energy $U$ used in the DFT calculations is in the range of 0 – 4 eV [20,40,44]. Different DOS was reported with different functional used in the calculations [41]. The problem has been that there is no high quality, direct experimental results available for the comparison with the calculated DOS. Here, by utilizing scanning tunneling microscopy and spectroscopy (STM/S), the electronic DOS of the exfoliated $CrBr_3$ flakes are revealed. The observed d$I$/d$V$ spectra agree well with all the reported optical measurements and the energy gap of bulk $CrBr_3$ can be determined to be 0.57 eV $\pm$ 0.04 eV, indicating $CrBr_3$ is a small gap semiconductor. With this data, a set of parameters for DFT calculations is suggested.

**Experiments and Methods**

The magnetization measurement was carried out with a Quantum Design Physical Property Measurement System (PPMS) on a single crystal $CrBr_3$. The sample was first cooled down to 2 K at zero field and magnetization *versus* temperature measurement was taken by increasing the temperature from 2 K with an in-plane, DC magnetic field of 500 Oe. $CrBr_3$ has a layered structure formed by van der Waals force with high in-plane stiffness making it easier to exfoliate into 2D layers [41]. In this study, the "scotch-tape" method [1] was used to directly exfoliate the $CrBr_3$ flakes (single crystals $CrBr_3$ bought from HQ graphene) onto a highly oriented pyrolytic graphite (HOPG) or Si/SiO$_2$ substrates [1]. First, a blue industrial tape was attached onto the top surface of a $CrBr_3$ crystal. When the blue tape was removed from the crystal, many $CrBr_3$ flakes were exfoliated from the crystal. A polydimethylsiloxane (PMDS) stamp was then in contact with the blue tape for the second exfoliation. Finally, the $CrBr_3$ flakes were transferred onto the desired substrate. It is worth noting that a slow peeling rate of PDMS could get more flakes transferred onto the target substrate. HOPG substrate is used for STM measurements, while Si/SiO2 wafer is used as substrate for scanning electron microscopy (SEM) and energy dispersive X-ray spectroscopy (EDS) measurements.

This exfoliation process was done in an ambient environment and the sample was transferred to the ultra-high vacuum (UHV) chamber within ~15 minutes of the air exposure. No annealing procedure was carried out after loading CrBr$_3$/HOPG into UHV prior to STM measurements to avoid any possible evaporation of the atoms that may create the point defects. The STM measurements were conducted with a base pressure of $10^{-11}$ mbar. d$I$/d$V$ signals were recorded with 997 Hz modulation frequency, 1 ms time constant, and 50 meV voltage modulation in the lock-in settings. STM topography and d$I$/d$V$ images were collected with 450 × 450 pixels.

The calculations were performed using projected augmented wave (PAW) method as implemented in the Vienna *ab initio* simulation package (VASP). In the calculations GGA and LDA pseudopotentials (PP) were used. All combinations involving the effective on-site exchange interaction, $J$, and the effective on-site Coulomb repulsion energy, $U$, with $J$ = 0, 1, 2, 3 and $U$ = 0, 1, 2, 4, 6, 8 were calculated for each PP. LDA+U calculations were performed by adding the interaction to the Cr $d$-orbital. The results from the GGA PP were rejected as they did not show zero energy gap when $U = J = 0$, expected in a Mott insulator. On the other hand, results from LDA PP exhibit a Mott insulator behavior: zero energy gap when both $U$ and $J$ are set to 0. A combination of $U$ = 5 and $J$ = 3 was chosen as it fits the best to the experimental results. A mesh of 5×5×5 $k$ points generated by the scheme of Gamma point is used for a primitive cell calculation. LDA+U calculations were performed with 500 eV energy cut-off.

**Results and Discussions**

CrBr$_3$ has a layered structure formed by van der Waals force with high in-plane stiffness making it easier to be exfoliated into 2D layers [41]. The CrBr$_3$ flakes were exfoliated and transferred onto highly oriented pyrolytic graphite (HOPG) substrates. Figure 1(a) illustrates the exfoliation procedures. Various flakes with different thicknesses are observed with different colors due to the light interference, as shown in Fig. 1(b). The magnetization vs temperature ($M-T$) measurement, as shown in Fig. 1(c), exhibits a clear Curie-Weiss behavior at high temperatures and a transition temperature of 31.5 K, consistent with previously work ($T_C$ = 32.5 K [45]). When the applied in-plane magnetic field is small, the magnetization decreases with the decreasing temperature below $T_C$ due to magneto crystalline anisotropy [46]. Figure 1(d) shows a large scale STM topography image taken from a thick (bulk) $CrBr_3$ where the step features with a height change of ~5.9 Å (single layer thickness of $CrBr_3$ [46]) are clearly visible.

Atomic resolution images are revealed with zoom-in STM topography images measured at 77 K, as shown in Fig. 2(b). This atomic resolution image was taken on a thick (bulk) $CrBr_3$ flake. Though the images are noisy, the atomic lattice structures can be resolved. The noisy measurement is a result of the combination of the poor conductivity of the $CrBr_3$ flakes and the fact that the samples were not degassed after loading to the UHV chamber. The purpose of not performing the degassing procedure under UHV condition is to prevent from creating Br defects. The tip condition was confirmed to be good with the atomic resolution images on nearby HOPG (the substrate) regions, as shown in Fig. S2. The atomic resolution image of the $CrBr_3$ agrees well with the crystal structure (space group $R\bar{3}$), as shown in Fig. 2(b). Note that the overlaid crystal structure model is a bilayer structure of $CrBr_3$, indicating the topographic images are sensitive to the second layer with the selected scanning parameters. This observation is different from a recently reported molecular beam epitaxy (MBE) grown $CrBr_3$ [9], in which only the topmost layer structure dominates the STM topographic images. The differences might be due to the different imaging conditions or different sample preparation methods. In this work, the $CrBr_3$ flakes were exfoliated and transferred onto the HOPG; while Chen *et al.* prepared their samples by MBE onto HOPG [9]. It is possible the MBE growing process may induce more CrBr$_3$-substrate interactions and result in different electronic properties compared to the exfoliated CrBr$_3$. On the other hand, the scanning set points used in this work (1 V; 1 nA, leads to R$_{junction}$ = 1 GΩ) make our tip-sample distance closer compare to that in the Ref. [9] (1.5 V; 0.5 nA, leads to R$_{junction}$ = 3 GΩ). The closer tip-sample distance may allow us to observe the second layer, but not with the scanning conditions used in Ref. [9]. As will be discussed later, the second layer sensitivity is also observed in the d$I$/d$V$ spectra of ML CrBr$_3$, shown in Fig. 4(c). More STM studies with varying tunneling conditions on CrBr$_3$ prepared with different methods are needed to clarify the origin of the discrepancy.

Furthermore, the normalized d$I$/d$V$ spectra taken at $T$ = 77 K with different bias ranges (± 0.8 V, ± 1 V, ± 2 V and ± 3 V with higher sensitivity setting in lock-in amplifier for smaller bias range measurements in order to observe smaller signal near Fermi energy) are shown in Fig. 2(c); while the raw d$I$/d$V$ spectra are shown in the supplementary document (Fig. S3(a)). Note that in the case of the spectra measured with ±3V bias range, peaks closer to Fermi level cannot be resolved due to the low signal and lower sensitivity setting. By comparing the peak intensity ratios, one can find that peak 4 has the intensity that is 43 times of that of peak 1. In the sensitivity setting for bias range of ±3V, when the peak 4 is visualized, the peak 1 would be in the level of the noise, hence, invisible. The multi-peak features are observed and confirmed with the derivative of the measured $I$-$V$ curves shown in Fig. S3(c). The normalized $dI/dV$ is done by $(dI/dV)/(\overline{I/V})$, where $\overline{I/V}$ is defined as [47]:

$$\overline{I/V} = \int_{-\infty}^{\infty} [I(V')/V'] e^{-\frac{|V'-V|}{\Delta V}} dV' \qquad (1)$$

where $\Delta V$ is the broadening width. To reduce the noise in the gap region, value of $\Delta V$ should be of the order of material band gap [48]. Here, values of $\Delta V$ are chosen to be $\pm 0.15$ V, $\pm 0.25$ V, $\pm 0.35$ V and $\pm 0.9$ V for spectra with bias ranges of $\pm 0.8$ V, $\pm 1$ V, $\pm 2$ V and $\pm 3$ V, respectively. These values are in the order of material band gap and are chosen to reduce the noises. Higher value of $\Delta V$ is needed to reduce larger noise in the spectra with wider bias range. Comparison between normalized (Fig. 2(c)) and raw spectra (Fig. S3(a)) shows the successful noise reduction without influencing the peak positions.

It is believed that the normalized d$I$/d$V$ spectra are proportional to the local density of states (LDOS) [47]. As shown in Fig. 2(c), the nine observed peaks are labeled with numbers, *1-4*, and letters, *a-e*, for peaks in the conduction and valence bands, respectively. The peak positions are analyzed, through Gaussian function fitting, from a total of eight measured d$I$/d$V$ spectra taken at randomly picked locations on the thick layer of exfoliated $CrBr_3$. All the peak positions are highly reproducible, regardless of the measurement locations. The results are shown in Fig. S3(b) and summarized in Table 1. We notice that the peak positions among different d$I$/d$V$ measurements occasionally exhibit a small peak shift, usually within 0.1 V. The observed small peak shifts are considered in the uncertainty shown in Table 1 and the bin width is chosen to be 0.1 V in Fig. S3(b). In order to compare with reported optical transitions, Table S1 shows the energy differences of all possible conduction-valence band peak pairs. It is worth noting that the d$I$/d$V$ spectra show two peaks nearest to the Fermi energy, one in the conduction band and the other in the valence band (peak "*1*" ($0.25 \pm 0.03$ eV) and peak "*a*" ($-0.32 \pm 0.02$ eV)). These two peaks determine the peak-to-peak energy gap of $CrBr_3$ to be $0.57 \pm 0.04$ eV, which is much smaller than previously believed values (1.68 – 2.1 eV [16,17,24,38]). The energy gap determined by the onset of the d$I$/d$V$ signal is $0.29 \pm 0.05$ eV, as indicated in Fig. 2(c). In addition, an observed PL peak (1.35 eV) was argued to be related to Cr *d-d* transition. This transition is not necessary to be related to the energy gap, it is matched to the peak pair *2-a* in our data here. One may argue that either peak *1* or *a* is caused by defects and not intrinsic. If it is the case, then the energy gap will be determined by either *1-b* (1.08 ± 0.05 eV), *2-a* (1.27 ± 0.04 eV) or *2-b* (1.78 ± 0.06 eV). We can rule out the possibility that either peak *1* or *a* is caused by the defects with the following facts: (1) the analyzed d$I$/d$V$ spectra are highly reproducible even they are measured at different locations on the crystal; and (2) the atomic resolution shown in Fig. 2(b) shows defect-free images. It is known that STM is a very local measurement, and the defect induced state will only be picked up by measurements performed near the defects [49]. With the facts mentioned above, we can unambiguously rule out that the peaks *1* and *a* are caused by defects. In other words, they are intrinsic.

The observed multi-peak feature in the d$I$/d$V$ spectra is found to match well with the previously reported optical transitions. Table 2 lists the reported low temperature optical measurements, including PL [26], absorption [14,15,17,23,24] Kerr rotation [14,31], reflection [25], and dielectric measurements [14]. The corresponding conduction-valence band peak pairs from Table S1 is also listed in Table 2. We find that all the reported optical transition energies (Table 2) are matched very well with energy differences between certain measured d$I$/d$V$ conduction-valence band peak pairs (Table S1). The extremely good agreements between the reported optical transition energies and the measured d$I$/d$V$ conduction-valence band peak pairs indicate that the measured d$I$/d$V$ is closely representing the CrBr$_3$ electronic DOS. This further confirms that the two small peaks near zero bias (peak *1* and *a*) are intrinsic. Note that most of the reported optical measurements were performed at low temperature (lower than the Curie temperature, $T_c$ = 32 K). The good agreements between the reported optical measurements with the d$I$/d$V$ spectra presented here indicate that the DOS of CrBr$_3$ does not change significantly with the magnetic phase transition.

The nature of the DOS is further investigated with DFT + U calculations. Figure 3(a) compares the measured d$I$/d$V$ spectrum (top panel) and the DOS calculated by DFT+U for (middle panel) $U = 5$ and $J = 3$; and for (bottom panel) $U = J = 0$. For the calculation with both $U$ and $J$ are 0, a gapless DOS is revealed. An energy gap is opened when $U$ and $J$ are non-zero. By varying a wide range of $J$ and $U$ values (J varied from 0 to 3 eV while U varied from 0 to 11 eV), it is found that the energy gap in the DOS calculated with $J = 3$ and $U = 5$ agrees reasonably well with the measured d$I$/d$V$ spectrum (Fig. 3(a), top panel). We note that DFT is a powerful method to study the physical properties, but it is based on mean field approximation with adjustable parameters to include the electron correlation effects. Thus, we do not expect that all the detailed features in DOS can be reproduced but only focused on the energy gap. Figure 3(b) shows the partial DOS contributed from the Cr-$d$ orbitals, Br-$p$ orbitals and with spin-up and spin-down configurations. This shows that the Br-$p$ dominates the valance band while the Cr-$d$ dominates the conduction band.

Next, let us move our attention to the monolayer (ML) CrBr$_3$ on the top of the HOPG substrate. Figures 4 (a) and (b) show the topography and d$I$/d$V$ mapping of a ML CrBr$_3$ measured at 77 K, respectively. The clear d$I$/d$V$ contrast (Fig. 4(b)) unambiguously distinguishes the substrate HOPG and the ML CrBr$_3$ flake. The d$I$/d$V$ spectrum (Fig. 4(c)) measured on the ML CrBr$_3$ exhibits a similar shape as the d$I$/d$V$ spectrum measured on HOPG (Fig. S2(d)) with additional humps in the positive bias matching the measured peaks on the bulk CrBr$_3$ d$I$/d$V$ spectrum (Fig. 2(c)). We believe that the combined spectral features are due to that the scanning condition is sensitive to the second layer material - HOPG. Evidently, the measured d$I$/d$V$ spectrum of the ML CrBr$_3$ placed on HOPG is consistent with a recent theoretical calculation with ML CrBr$_3$ placed on graphene [50]. However, with the influences of HOPG on the measured d$I$/d$V$ spectrum, it is impossible for us to extract the energy gap information on the ML CrBr$_3$. Whether if the energy gap of ML CrBr$_3$ may be different from that of the bulk CrBr$_3$ is still unclear. Further experiments are needed to provide the insights on this question.

The monolayer thickness is confirmed and measured by the line profile across the substrate (HOPG) and the monolayer flake (Fig. 4(d)) as ~4.8 Å, which is smaller than the step height (5.9 Å) found in the bulk region (Fig. 1(d)). This smaller apparent height is most likely due to the electronic contrast between the CrBr$_3$ and HOPG. As the HOPG is more conductive than CrBr$_3$, under the constant current scanning mode, the tip-sample distance will be smaller for the tip-CrBr$_3$ case compared to that of the tip-HOPG case, which leads to ~1 Å lower apparent height compared to that measured in the bulk CrBr$_3$. On the ML CrBr$_3$ flake, there are a few small islands, which are bilayer (BL) CrBr$_3$, judged by the similar d$I$/d$V$ contrast and the corresponding height change in the topography. The height of the second layer on top of the ML CrBr$_3$ is measured to be 3.8 Å (Fig. 4(e)), which is even smaller than the monolayer case (4.8 Å). We believe this is because of the incomplete CrBr$_3$ second layer due to the edge degradation and the small island size. The round shape, instead of flat top shape, topographic profile (Fig. 4(e)) is a clear evidence. The edge degradation, due to the short air exposure (~15 min) prior to loading into the UHV chamber, is also clearly seen in Fig. 4(b). The edge of the monolayer is curved and exhibits high d$I$/d$V$ contrast with uniform length scale of ~2.3 nm. Similar degradation is also observed at the edge of the second layer with ~1.5 nm lateral length scale (Fig. 4(e)). The smaller lateral length scale of the degradation in the second layer indicates either (i) the edge in the second layer is relatively more stable than that of the monolayer edge; or (ii) the smaller lateral sizes of the second layer islands measured here limit the degradation process.

**Conclusions**

Using STM/S, atomic resolution image and electronic properties of $CrBr_3$ flakes have been explored. A small energy gap of $CrBr_3$ measured at 77 K is observed as the peak-to-peak energy gap of $0.57 \pm 0.04$ eV and the d$I$/d$V$ signal onset energy gap of $0.29 \pm 0.05$ eV. It is argued that the DOS is not sensitive to the magnetic phase transition ($T_C$ = 32 K). The atomic resolution image indicates that the observed small energy gap is not due to defects. Excellent agreements with reported optical transition energies further confirm that the peak features in the measured d$I$/d$V$ spectra are intrinsic to $CrBr_3$. The DFT calculation along with the observed d$I$/d$V$ spectra confirms that $CrBr_3$ has a smaller energy gap than ever reported. The ML and BL $CrBr_3$ measurements further provide information of the edge degradation due to the short air exposure. Our results provide a complete understanding of the electronic properties of the $CrBr_3$, including the reveal of the small energy gap.

**Acknowledgment**

D.B. acknowledges the Academic Affairs Energy GA Fellowship from University of Wyoming. J. Tang, J. Tian, Y.D., and T.Y.C. acknowledge the U.S. Department of Energy, Office of Basic Energy Sciences, Division of Materials Sciences and Engineering for financial support (DE-SC0020074) of this research. We thank Dr. Michael A. McGuire and Prof. Hua Chen for valuable discussions.

**Author Contributions**

D.B., R.D., A.W. performed STM measurements. Z.F. performed EDS. A.S.Z. calculated DOS from DFT calculations. N.S. and U.E. provide PPMS measurements. T.Y.C., J. Tian, Y.D. and J. Tang supervised the project. All authors contributed to the manuscript writing and discussion for the conclusions.

**Competing Interests statement**

The authors declare no competing interests.

**Figures & Legends**

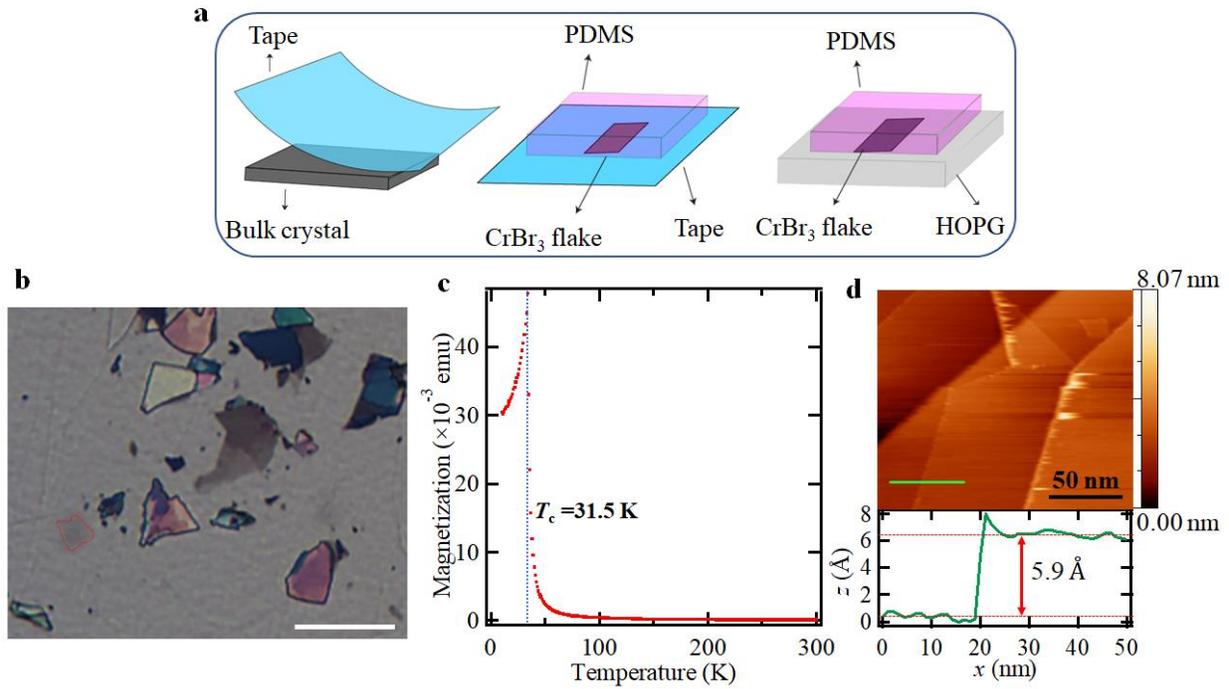

**Figure 1. a,** Schematic of "scotch-tape" technique used to exfoliate CrBr$_3$ flakes onto a HOPG substrate. **b,** Optical image of transferred CrBr$_3$ flakes on a HOPG substrate (scale 50 μm). Light gray regions have thinner layers (≤ 3 ML; one example- enclosed with red boundary), bright regions have thicker layers (>3 ML) of CrBr$_3$. **c,** Magnetization ($M$) *vs* temperature ($T$) measurement showing a magnetic transition temperature of 31.5 K. **d,** STM topography image of thick (bulk) CrBr$_3$ flake at room temperature (Bias: 2 V, set point current: 400 pA). The height profile indicates that the step height is corresponding to the monolayer thickness.

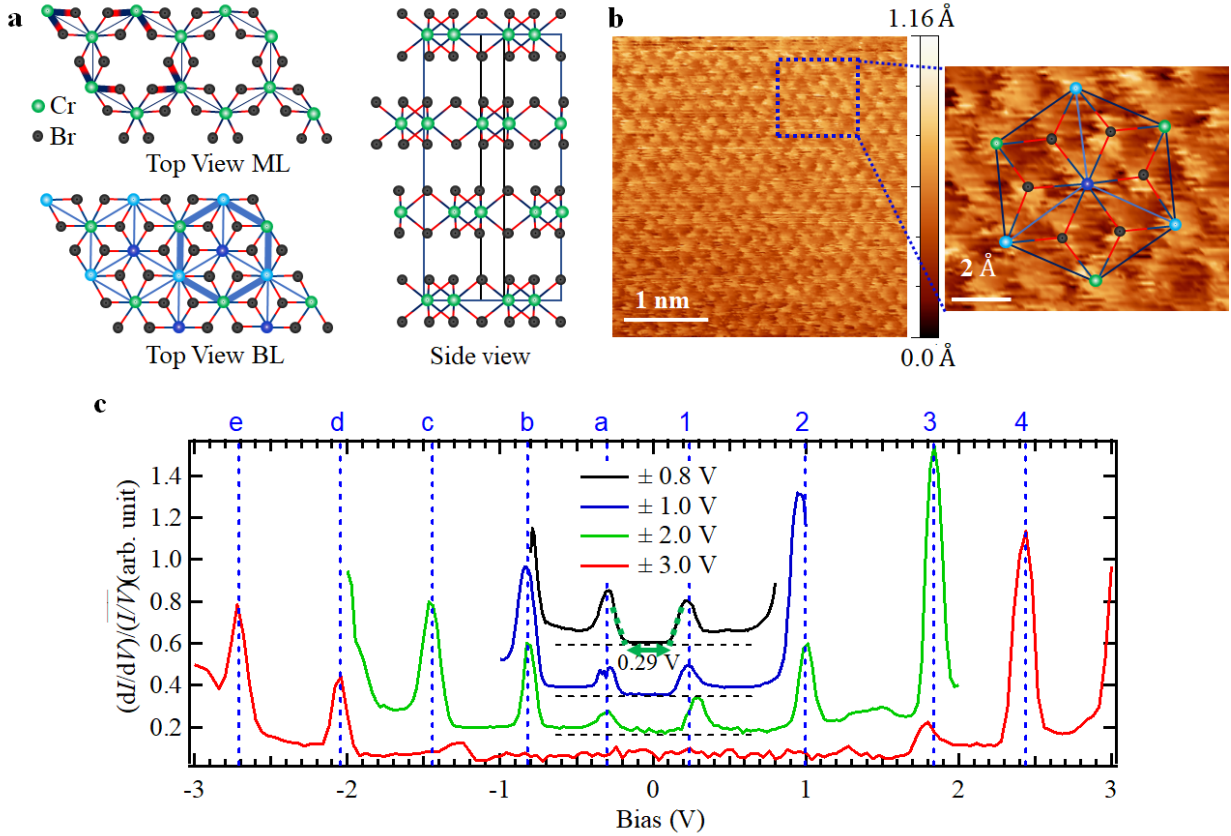

**Figure 2. a,** The top view of the crystal structure of $CrBr_3$ (ML: monolayer, BL: bilayer). In the ML structure, the unit cell is indicated with the thick bonds. In the BL structure, the green circles represent Cr atom at the top layer, the blue circles represent Cr at the bottom layer and the sky-blue circles represent Cr atoms appeared at both layers. The thick blue hexagon shows the overlaid structure used in b (the magnified image). **b,** Atomic resolution STM image of thick (bulk) $CrBr_3$ (Bias: 1 V, tunneling current: 1 nA). Magnified view of the region marked by blue dashed box overlaid with BL crystal structure of $CrBr_3$. **c,** Normalized $dI/dV$ spectra of $CrBr_3$ (bulk) taken at 77 K with different bias ranges and sensitivity settings in the lock-in amplifier. The $dI/dV$ spectra are shifted vertically for easy comparison. Horizontal dashed lines (black) represent the zero-base line of each spectrum. The set point conditions used prior to the $dI/dV$ measurements are (i) 0.8 V, 1 nA for $\pm$ 0.8 V; (ii) 1 V, 1 nA for $\pm$ 1 V; (iii) 2 V, 1 nA for $\pm$ 2V; and (iv) 3 V, 400 pA for $\pm$ 3 V. green double headed arrow shows the onset band gap.

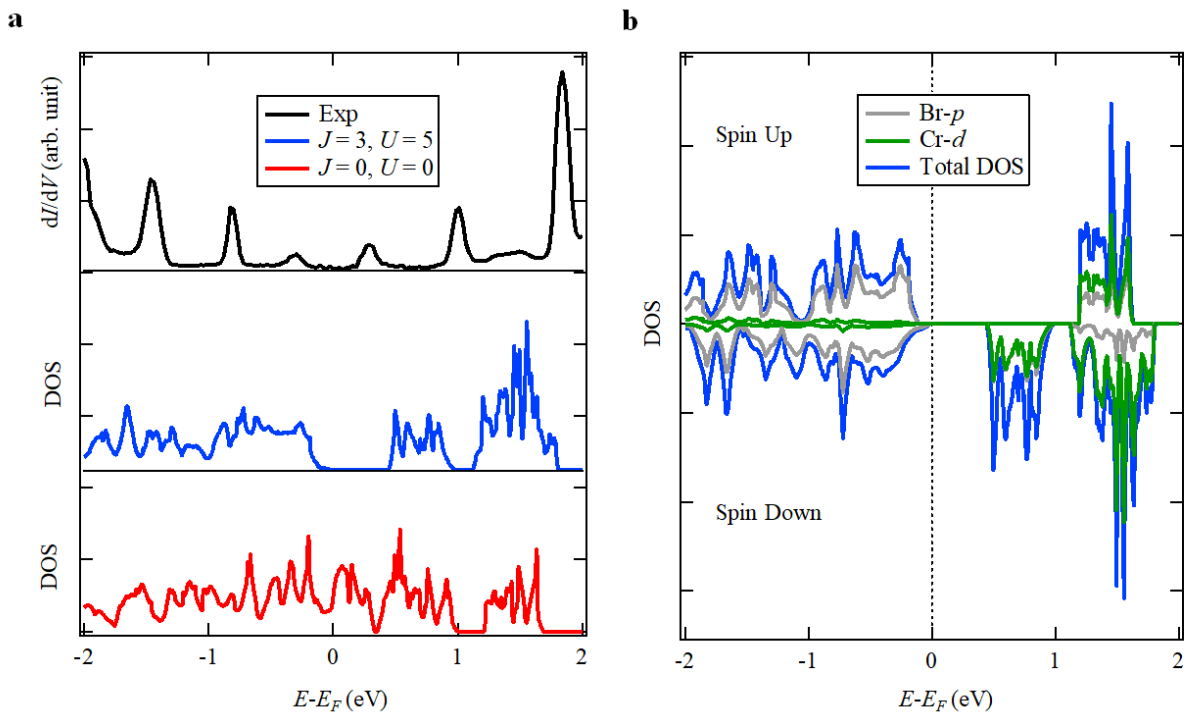

**Figure 3. a,** Comparison of DFT + U calculation (bottom panel: $J = 0$ eV, $U = 0$ eV; and middle panel: $J = 3$ eV, $U = 5$ eV) with the d$I$/d$V$ spectra (top panel). **b,** Partial DOS from Cr $d$ orbitals (green), Br $p$ orbitals (gray) showing the orbital contribution to the total DOS (blue). Upper and lower panels of the graph shows the contributions from the spin-up and spin-down states.

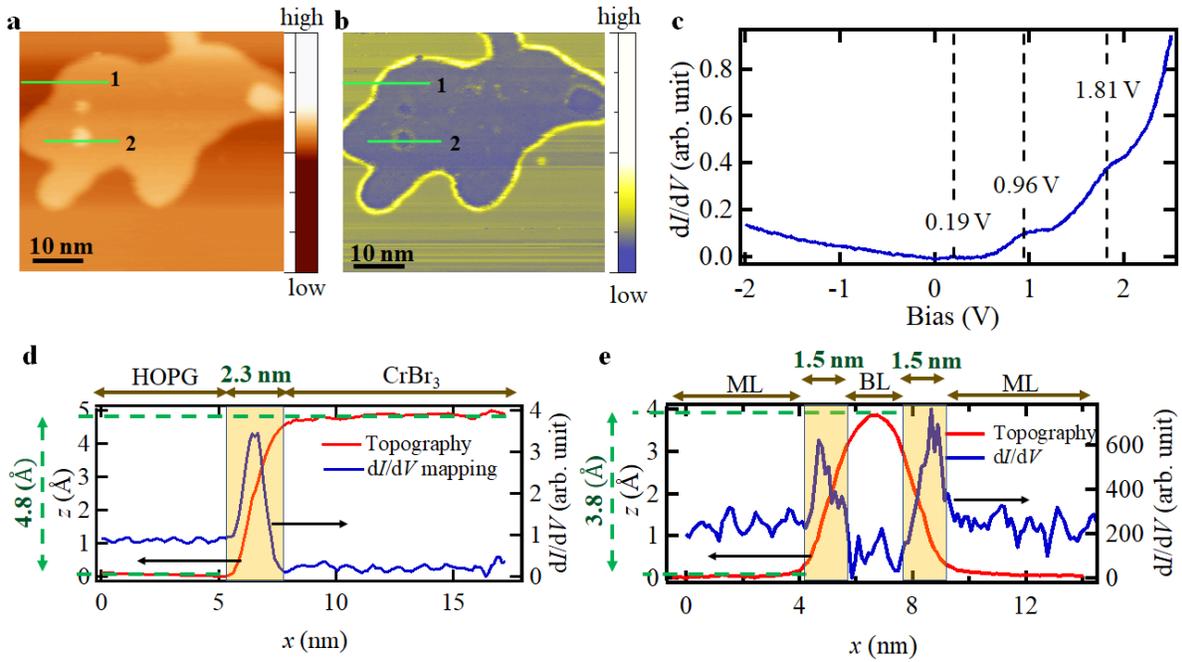

**Figure 4. Monolayer and bilayer CrBr$_3$ topography and d$I$/d$V$ measured at 77 K.: a,** Topography and **b,** d$I$/d$V$ images of the monolayer CrBr$_3$ on HOPG and small second layer islands on top of the monolayer CrBr$_3$. Imaging condition: bias = 4 V; set point current = 400 pA. **c,** d$I$/d$V$ spectrum of the monolayer CrBr$_3$ on HOPG. Set point condition prior to the d$I$/d$V$ measurement is 4 V, 400 pA. **d** and **e,** Line profiles of the topography and d$I$/d$V$ mappings across the HOPG/monolayer (line 1) and monolayer/bilayer/monolayer (line 2), as indicated with green lines in (a) and (b).

## Tables

**Table 1.** Peak positions of the labeled peaks in Fig. 2(c).

| Assigned peak symbols | Peak positions (eV) | |
|---|---|---|
| 4 | 2.38 ± 0.08 | |
| 3 | 1.82 ± 0.04 | Conduction band |
| 2 | 0.95 ± 0.05 | |
| 1 | 0.25 ± 0.04 | |
| a | − 0.31 ± 0.02 | |
| b | − 0.83 ± 0.05 | |
| c | − 1.43 ± 0.02 | Valance band |
| d | − 2.06 ± 0.01 | |
| e | − 2.72 ± 0.01 | |

**Table 2.** Comparison between the energies of the reported optical transitions and the conduction-valence peak pairs.

| Measurement | Measurement temperature | Assigned transition | Reported energy gap | Corresponding energy gap (eV) | Peak pair | Peak pair energy (eV) | Ref. |
|---|---|---|---|---|---|---|---|
| PL | 2.7 K | | 1.35 eV | 1.35 | 2 – a | 1.27 ± 0.05 | [26] |
| Abs. | 4.2 K | $^4T_2$ | 13500 cm$^{-1}$ | 1.68 | 1 – c | 1.68 ± 0.04 | [23,24] |
| Abs. | 4.2 K | $^2T_1$ | 14400 cm$^{-1}$ | 1.79 | 2 – b | 1.78 ± 0.07 | [23,24] |
| Abs. | 4.2 K | $^4T_1$ | 17500 cm$^{-1}$ | 2.17 | 3 – a | 2.14 ± 0.04 | [23,24] |
| Abs. | 4.2 K | $^2T_2$ | 18900 cm$^{-1}$ | 2.36 | 2 – c or 1 – d | 2.38 ± 0.05 or 2.31 ± 0.04 | [24] |
| Abs. | 4.2 K | | 19200 cm$^{-1}$ | 2.38 | | | [23] |
| Kerr (+ max) | 1.5 K | $t_{1u}{}^n \rightarrow e_g{}^*$ | 23500 cm$^{-1}$ | 2.92 | 1 – e or 2 – d | 2.97 ± 0.04 or 3.01 ± 0.05 | [14,15,31] |
| Abs. | 1.5 K | $t_{1u} \rightarrow e_g{}^*$ | 24500 cm$^{-1}$ | 3.04 | | | [15] |
| Abs. | 1.5 K | | 24310 cm$^{-1}$ | 3.02 | | | [14] |
| Refl. | 30 K | | 3.1 eV | 3.1 | | | [25] |
| Dielectric | 1.5 K | | 25000 cm$^{-1}$ | 3.1 | | | [14] |
| Kerr (− max) | 1.5 K | $t_{2u}^n \rightarrow e_g^*$ | 26700 cm$^{-1}$ | 3.29 | 3 − c or 4 − b | 3.25 ± 0.04 or 3.21 ± 0.09 | [14,31] |
| Abs. | 1.5 K | | 29500 cm$^{-1}$ | 3.66 | 2 − e | 3.67 ± 0.05 | [15] |
| Abs. | 1.5 K | | 29580 cm$^{-1}$ | 3.67 | | | [14] |
| Refl. | 30 K | | 3.8 eV | 3.8 | 4 − c or 3 − d | 3.81 ± 0.08 or 3.88 ± 0.04 | [25] |